\documentclass[11pt,twoside]{article}
\usepackage{asp2004}
\usepackage{psfig}
\usepackage{epsf}
\usepackage{graphics}
\usepackage{lscape}
\usepackage{natbib}
\def\edcomment#1{\iffalse\marginpar{\raggedright\sl#1\/}\else\relax\fi}
\reversemarginpar

\def\etal{{et~al. \,}}

\begin{document}
\title{Rotating Core Collapse and Bipolar Supernova Explosions}
\author{A. Burrows}
\affil{Department of Astronomy, The University of Arizona, Tucson, AZ 85721;
aburrows@as.arizona.edu}
\author{R. Walder}
\affil{Department of Astronomy, The University of Arizona, Tucson, AZ 85721;
rwalder@as.arizona.edu}
\author{C.D. Ott}
\affil{Max-Planck-Institut f\"{u}r Gravitationsphysik,
Albert-Einstein-Institut, Golm/Potsdam, Germany; cott@aei.mpg.de}
\author{E. Livne}
\affil{Racah Institute of Physics, The Hebrew University,
Jerusalem, Israel; eli@frodo.fiz.huji.ac.il}

\begin{abstract}

We review some of the reasons for believing that the generic
core-collapse supernova is neutrino-driven, not MHD-jet driven.
We include a discussion of the possible role of rotation in
supernova blast energetics and morphology, and speculate on the origin
of Cas A's and SN1987A's ejecta fields.  Two ``explosive" phenomena
may be associated with most core collapses, the neutrino-driven supernova
itself and an underenergetic jet-like ejection that follows.  
The latter may be a magnetic wind that easily penetrates the debris 
created by the much more energetic supernova.  We predict
that many core-collapse supernova remnants should have
sub-dominant jet-like features.  In Cas A, 
we associate this sub-dominant collimated wind with the ``jet/counter-jet"
structure observed.  We suggest that the actual Cas A explosion itself
is at nearly right angles to this jet, along a rotation axis that coincides
with the bulk of the ejecta, the iron lobes, and the putative direction
of motion of the point source.  It may be that when rotation becomes sufficiently
rapid that the strong-neutrino-driven-supernova/weak-jet duality
switches to a strong-MHD-jet scenario that might be associated with
hypernovae, and in some cases GRBs. Finally, we present a calculation
using a new 2D multi-group, flux-limited radiation/hydrodynamics code we have
recently developed for the simulation of core-collapse supernovae. 
We discuss the rotation-induced anisotropy in the neutrino radiation 
field, neutrino heating, and the neutrino flux vectors 
and speculate on rotation's possible role in  
the supernova mechanism and in the overall supernova phenomenon.

\end{abstract}
\thispagestyle{plain}

\section{Neutrino-Driven Mechanism, Perhaps with Rotation, Preferred Over MHD Jet-Driven Mechanism}
\label{intro}

We do not yet have an agreed-upon explanation for the mechanism of 
core-collapse supernovae, nor for the related phenomena of 
neutron star kicks and blast debris morphology.
Nevertheless, there has recently been significant progress in understanding 
the multi-dimensional hydrodynamics and neutrino
radiative transfer of implosion, bounce, and explosion and a picture is emerging, however dim,
of what can and can not be true.  It has been suggested that  
the association between gamma-ray bursts (GRBs), which seem to be
powered by relativistic jets, and the Type Ic subclass of supernovae,
as well as the existence of hypernovae \citep{mazzali}, means that 
the generic supernova is powered my MHD jets \citep{hoflich,akiyama}.  However,
there are many reasons to doubt this conclusion.
The GRB rate in the universe is close to one or two per day,
but the universe's supernova rate from the death of stars more massive than
$\sim$8 M$_{\odot}$ is close to one per second.  With a beaming
angle of $\sim$5$^{\circ}$ ($1/500$th of the sky), the GRB/SNe ratio
is still only $\sim$0.5\%.  This is significantly lower than the fraction
of all supernovae that are Type Ic's.   
In addition, the spectra of canonical Type Ic supernovae 
do not show the high velocities and high $^{56}$Ni masses
associated with the bumps in GRB afterglows and SN1998bw-like explosions \citep{mazzali,maeda}.  
This implies that the supernovae paired with GRBs are special \citep{pod}.
Hence, though evidence is accumulating that GRBs are 
connected with the deaths of a subset of massive stars, that subset
is likely to be small.  For instance, a major contender for the GRB engine, the collapsar model \citep{macfadyen},  
could involve only those stars more massive than 80 M$_{\odot}$ that also have ``rapidly"
rotating cores at collapse. The massive star mass function places
$\sim$1-2\% of all massive stars above this cutoff, still enough to explain
GRBs even if only a minority of them are fast rotators.

Nevertheless, it has been argued that ``MHD" jets power 
both supernovae and GRBs, but that the continuum of progenitor 
masses translates into a continuum of explosion energies, gamma-ray bursts
being at the more violent and relativistic end.  
Evidence is emerging, however, that supernova energies and GRB energies
are comparable, both ranging from $\sim$10$^{50}$ ergs to some multiple of 10$^{51}$ ergs.
There is an outside chance that the average GRB is 
actually less energetic than the average supernova.
Moreover, it has been suggested that the only major difference between supernovae 
and GRBs is the presence or absence of hydrogen and helium envelopes. 
But, as already mentioned, the typical Type Ib/c supernova
light curves, line profiles, spectra, and ejection velocities 
look very different from those inferred by analyzing hypernovae and GRB afterglow bumps.
Furthermore, the majority Type IIp supernovae reveal no morphological 
or spectral signatures of envelope-decelerated jets.

It has been claimed that the large polarizations seen in
many Type Ic supernovae \citep{wang2} are evidence that the necessary asymmetry is due
to a jet morphology.  However, any explosion asymmetry, such as has been
a feature of supernova theory for the last decade 
\citep{bhf,fryer2002,rampp20022,buras2003,scheck}, could explain
the magnitude of the measured polarizations.  The 
instabilities that are generic to neutrino-driven models
produce large-scale fingering and asphericities with 1:2 or 1:3 axis ratios \citep{bhf}.
However, the angle of polarization observed in SN1987A is roughly correlated
with the axes of its encircling rings and these rings are likely associated
with a rotation axis.  Therefore, it is reasonable to conclude that 
SN1987A's explosion asymmetry, as inferred from polarization studies, 
is correlated with the rotation of the core.  In addition, SN1987A's debris 
morphology is bipolar, and is along the same axis \citep{wang1}. However,
this does not mean that supernovae are driven by jets.  In fact, introducing
rotation into the standard neutrino-driven paradigm of
core-collapse supernovae can produce quasi-bipolar explosions of the requisite
character (\S\ref{rotation}; Fryer \& Heger 2000; Burrows, Ott, \& Meakin 2003).   The 
bipolarity is not strongly collimated (30-60$^{\circ}$), unless
the rotation rate is large \citep{bom}.  Note that, depending upon 
geometrical details and stellar profile, some of the shock of a bipolar explosion
can meet itself (``clap") at the equator, producing an equatorial spray.
Such a spray, a bit off-center, may be responsible for the fact that
the ring around SN1987A lights up on one side before the other and that
its manifest equatorial speeds are higher than those of the 
inner visible quasi-bipolar debris. 

With rapid rotation, MHD power could be
dynamically influential, but this may not be the normal situation.     
A picture we favor is that most supernovae are neutrino-driven, with
the aspherical blast morphology influenced by slow to modest rotation.
This will give a bipolar structure to the explosion debris and polarize
the inner ejecta.  It will also result in a bipolar distribution
of the iron-group elements.   Such a morphology and
element distribution are seen in Cas A (\S\ref{casa}; 
Hwang et al. 2004).  In the neutrino-driven scenario, 
not only does rotation impose modest bipolarity, but it enhances the 
chances for explosion and increases the explosion energy (\S\ref{rotation}).
Moreover, the neutrino-driven supernova may be accompanied by 
a sub-dominant MHD jet with a low luminosity and total energy, that may
nevertheless emerge from the core to modify the debris.  This jet may simply
be the expected protoneutron star wind that follows the supernova \citep{bhf}, modified
by rotation and magnetic hoop stresses.   Hence, there
may be two ``explosions" at different times: one, the neutrino-driven supernova
and, the other (a weaker one), the MHD jet or B-field-modified protoneutron star
wind.  For most supernovae, the latter two would be of minor
dynamical import.  

However, when the rotation rate is very fast, MHD processes could take over, particularly
if a massive core in a massive progenitor inhibits the neutrino-driven 
mechanism.  In that case, a hypernova and/or a GRB could result (if most of the outer envelope
has already been lost; perhaps also if a black hole first forms), driven
by a narrow ($\sim$3-10$^{\circ}$) MHD jet of high power.  
Rotation naturally amplifies
B-fields, either by the MRI (Akiyama et al. 2003; Thompson, Quataert, \& Burrows 2004 [TQB]), flux wrapping, or
more conventional dynamo action, and there will
be a critical rotation rate and degree of differential rotation
above which MHD power and power densities will take
over from those of neutrino processes.  But such
high spin rates may not be common, making hypernovae
and GRBs rare.  In addition, for suitably massive progenitors
with slowly rotating cores, the supernova may abort, a black hole may form,
and a hypernova/GRB may not follow.  This may be the normal evolutionary path for
most very massive stars ($> 40\ {\rm M}_{\odot}$??).  

In the context of the neutrino-driven supernova scenario, even when the
rotation rates are modest or low, there should be interesting
magnetic processes and B-field evolution.
Differential spin will amplify magnetic fields, 
perhaps to high values (10$^{13}$-10$^{15}$ gauss) that are either
transients and decay to more canonical pulsar fields of 10$^{12}$ gauss, or
survive to explain magnetars.  Neutrino-driven convection can 
drive dynamo action \citep{duncan}, will advectively redistribute magnetic
fields, and will alter the multipolarity structure.  In
addition to the dominant dipole, higher-order multipoles
on the surfaces of radio pulsars may be produced and bear the stamp
of early hydrodynamics.

\section{The Cas A Blast Morphology: Clues to the Supernova Engine}
\label{casa}

Cas A shows a jet in its northeast, and a ``counter-jet" in its southwest \citep{hwang04}
that some have suggested is a signature of a jet-driven explosion.  However,
most of the blast mass and energy is actually nearly perpendicular to this jet, concentrated
near the north and south poles \citep{will02,will03}.  Furthermore, iron that must be associated with
the central engine is distributed in two caps in the same locations \citep{hwang01,hwang04}.  We surmise
that the axis connecting these iron-rich, mass-rich, and energy-rich regions is the 
real axis of the supernova and that it exploded in quasi-bipolar fashion roughly along 
a rotation axis connecting the north-northwest/south-southeast directions.
The neutrino mechanism with some rotation can explain this morphology and debris
pattern (\S\ref{rotation}; \S\ref{aniso}).  The jet/counter-jet structures would not have driven the 
supernova, would be underenergetic, and would have emerged after the supernova 
explosion into the prepared supernova debris.  However, it would be natural 
for the jet axis to be a rotation axis, and since in Cas A this axis is roughly perpendicular
to the axis we are identifying with the bipolar explosion, the rotation axis
of the engine would have to have precessed between the explosion and the emergence 
of the ``MHD" jet or collimated protoneutron star wind. This is problematic, but, 
nevertheless, the rest of the story we outline for Cas A retains its appeal.
Alternately, the jet/counter-jet axis could indicate the direction of the 
magnetic dipole; a finite angle between the rotation axis and the magnetic dipole
axis is a central feature of pulsar theory.  If this were the case, there would
be added constraints on the ratio of the jet duration and the rotation period.
Note that the inferred direction of motion (kick?) of the central X-ray point source 
in Cas A \citep{thor} is along the axis we identify as the axis of explosion, not the 
``jet/counter-jet" axis.  This point source is also inferred
to be moving in the ``blue-shifted" direction, mildly towards us.  
In this scenario, the kick would be due to 
hydrodynamic recoil and (perhaps) gravitational attraction with the inner debris 
(H.-T. Janka, this conference; Scheck et al. 2004) in the context of a mildly top-bottom
asymmetric \citep{bh,scheck} bipolar explosion. Encouragingly,  
estimates are that the ``red-shifted" cap of mass and 
iron in the north/northwest has a bit more mass
and kinetic energy than the ``blue-shifted" cap, consistent 
with the recoil interpretation \citep{will02,will03}.  The gravitational 
impulses might be accompanied (in 3D) by gravitational torques that could 
precess the rotation axis of the settling nascent protoneutron star.

\section{The Neutrino-Driven Explosion Mechanism}
\label{mech}

The prompt bounce never leads to direct explosion in 1D, 2D, or 3D; neutrino losses and photodissociation
by the shock debilitate it, even for the lowest mass progenitors and accretion-induced
collapse (AIC).  In the Chandrasekhar context, there is just too much mass
between the place the shock originates ($\sim$0.6 M$_{\odot}$) 
and the outer boundary ($\ge$1.2 M$_{\odot}$) and 
the shock stalls into an accretion shock.  Furthermore, 
in spherical symmetry (1D), it has been shown using Boltzmann neutrino
transfer and the best physics that the delayed neutrino 
mechanism does not work either \citep{mezz2001,lieben2001,Tod1}.
In 1D, the bounce shock stalls and is not revived, though an increase of only $\sim$25\%
in neutrino heating would lead to explosion!  Such an increase could 
arise from as-yet-unknown neutrino effects or overturning motions 
in the inner core that could boost the neutrino luminosity. The former are unlikely and the latter 
have yet to be demonstrated. 

However, in 2D, but using less sophisticated
neutrino transfer (e.g., gray; 1D transport along multiple radial rays), 
numerous calculations result in explosions \citep{herant,bhf,fryer2002}, although sometimes weak.  These calculations
demonstrate that neutrino-driven convection in the so-called ``gain" region near the shock \citep{bethe}
increases the efficiency of neutrino energy deposition,  
increases the size of the gain region, and facilitates explosion.  
Figure \ref{fig:1} depicts the various
important regions.  It is neutrino
energy deposition in the gain region, not radiation pressure, that is important and if this deposition
is adequate explosions are easy.   The increase in efficiency in 2D
can be traced to the increase in the average dwell time ($\tau_{adv}$) of the matter in the gain region before
matter settles into the cooling region and onto the inner core.  If the heating time ($\tau_{H}$) 
is small compared with the hydrodynamic accretion/advection time into the interior ($\tau_{adv}$),
then the object explodes \citep{tqb}.  If $\tau_{H}$ is larger than $\tau_{adv}$, then the object does not
explode.   This qualitative condition is simple and multi-D effects increase $\tau_{adv}$. In
fact, recent 3D calculations (H.-T. Janka, this conference), still using 1D transport along radial rays,  
reveal that 3D is marginally better than 2D.  However, even this 3D explosion
is underenergetic by factors.  We speculate (\S\ref{rotation}) that 
a little rotation (a ``rotation boost") can make the 
difference in explosion energy and viability,
that rotation is the ``missing key," though how much and how it 
is or needs to be distributed has yet to be determined.  
Nevertheless, current calculations of the delayed neutrino mechanism 
have revealed that it is so close to working that we would be surprised
if it weren't essentially correct.  

Note that delay is a good thing, for it ensures that the remaining
neutron star has sufficient mass to explain measured pulsar masses and that 
the ejected material is not too neutron-rich to be inconsistent
with nucleosynthetic constraints.  Recent calculations \citep{tqb} 
demonstrate that using Boltzmann $\nu_e$ and $\bar{\nu}_e$ transport
in the delayed explosion context raises the Y$_e$ of ejected matter as it
emerges to values equal to or slightly greater than 0.5.   This does not happen
with gray or more approximate schemes \citep{bhf,fryer2002}, nor for rapid explosions with little delay.

\subsection{A Digression on the Basics of Supernova Energetics}
\label{energetics}

The supernova explosion is a phenomenon of the outer mantle at 
ten times the radius of a cold neutron star.
Though the binding energy of a cold neutron star is $\sim$$3 \times 10^{53}$ ergs
and the supernova explosion energy is near $10^{51}$ ergs, a comparison of these two energies and
the large ratio that results are not very relevant.  More relevant are the binding energy
of the mantle (interior to the shock or, perhaps, exterior to the neutrinospheres)
and the neutrino energy radiated during
the delayed phase.  These are both at most a few$\times$10$^{52}$ ergs, not
$\sim$$3 \times 10^{53}$ ergs, and the important ratio that illuminates
the neutrino-driven supernova phenomenon is $\sim$10$^{51}$
ergs divided by a few$\times$10$^{52}$ ergs. This is $\sim$5-10\%, not
the oft-quoted 1\%, a number which tends to overemphasize the
sensitivity of the neutrino mechanism to neutrino and numerical details.
Five to ten percent of the neutrino energy coursing through
the semi-transparent region is required, not one percent. Importantly, the optical
depth to neutrino absorption in the gain region is of order $\sim$0.1.  The product
of the sum of the $\nu_{e}$ and $\bar{\nu}_e$ neutrino energy emissions in the first
100's of milliseconds and this optical depth gives a number near 10$^{51}$ ergs.
Perhaps, this is not a coincidence.

\section{Core Rotation: The Missing Key?}
\label{rotation}

The amplification of the angular velocity due to large changes in the 
radius of a given mass shell during collapse implies
that rotation may be a factor in core collapse and in the
explosion mechanism. There are a few aspects of rotating 
collapse that distinguish it from
spherical collapse:  1) Rotation provides centrifugal support and lowers the effective
gravity in the core, increasing the radius of the stalled shock and the
size of the gain region.  Since ejection is inhibited by the deep
potential well, this consequence of rotation alone might facilitate explosion.
2) Rotation enhances the neutrino flux and heating rate along the poles, 
relative to those at the equator (Shimizu et al. 2001; \S\ref{aniso}),
perhaps facilitating a bipolar explosion (Burrows, Ott, \& Meakin 2003; 
Fryer and Heger 2000; Kotake, Yamada, \& Sato 2003).  
3) Due to the centrifugal barrier, rotation results
in an anisotropy in the mass accretion flux after bounce.  
If the rotation is ``rapid," funnel structures are 
generated at the poles (see Figure \ref{fig:2}; Burrows, Ott, \& Meakin 2003).
This, and the enhancement of neutrino energy deposition
along the poles (\S\ref{aniso}), suggest that the neutrino-driven mechanism could be 
along the poles.  Clearly, bipolarity is not an exclusive signature of MHD driven explosions
and may be a natural consequence of the neutrino-driven mechanism with rotation.
However, if the rotation rate is too low for these
polar effects to manifest themselves significantly, rotation might still lower
the effective gravity and mass accretion rate in the {\it equatorial} regions enough to lead to an 
equatorial, not a polar, explosion.  What obtains computational and physically 
has yet to be determined, but the potential variety rotation introduces 
should stimulate much research.  Finally,  
4) rotation generates vortices (Eddington-Sweet) that might dredge up heat from
below the neutrinospheres and thereby enhance the driving neutrino luminosities.

Of course, the actual magnitude of all these effects is a function of the initial rotational profile.
Nevertheless, the crucial questions that remain to be answered are: How much rotation is necessary to validate
the basic paradigm we have outlined?  What do we mean by ``rapid," ``modest," or slow rotation?  At
what critical rotation rate is the transition from neutrino-driven
to ``MHD"-driven?  ``Modest" rotation may not be as rapid as some think, but  
these questions are the subjects of current research. As noted, it is important
to know not only the spin rates, but the initial spin profiles.  
Models to date (see, e.g., Ott et al. 2004) have
generally employed a rotation law:

\begin{equation}
\label{eq:rotlaw}
\Omega(r) = \Omega_0 \, \bigg[ 1 + \bigg(\frac{r}{{\rm A}}\bigg)^2 \bigg]^{-1}\, ,
\end{equation}
where $\Omega(r)$ is the angular velocity, $r$ is the distance from
the rotation axis, and $\Omega_0$ and A are free parameters that
determine the rotational speed/energy of the model and the
distribution of angular momentum.  However, this law, for a given $\Omega_0$, puts a disproportionate
amount of spin kinetic energy in the core and can result, without
dissipative or moment-arm angular momentum transport after bounce, in
small final-state pulsar spin periods (2-6 milliseconds).  These might
be too fast to be generic.  True, through a large moment
arm or viscosity \citep{Tod2,tqb}, a transient high B-field phase might
transport angular momentum out of the core on dynamical times, leaving
protoneutron stars and pulsars with more subdued 10--50 millisecond ``birth" periods.
In fact, a high-field/high-spin state which had dynamical
consequences might, through B-field-aided spindown, erase its initial presence, at least
partially and/or some of the time.  Nevertheless, leaving behind pulsars
with ``rapid" spin rates is a concern.  However, the supernova involves
the envelope of the iron core and it is the initial spin rate out there which affects supernova
dynamics most.  Hence, one need not employ eq. (\ref{eq:rotlaw}).
One can start with a ``low" internal spin rate, which
would leave a lower pulsar spin rate, while still 
profiting on the outside from most of rotation's advantages (\S\ref{rotation}).
The lion's share of that angular momentum, if not shed by torques, would simply be ejected
in the explosion.  Due to partial centripetal breaking, a 
higher spin rate in the exterior would decrease the accretion
rate after bounce that may be inhibiting explosion.  It would also
lower the effective gravity which neutrino heating is ``trying" to overcome.
The question then would be: How realistic is such an 
initial rotational profile \citep{heger00,langer,hirschi}?

Even if the core is left with a high spin rate and we needed to
shed angular momentum via viscous torques \citep{tqb} to achieve the 
low inferred initial pulsar spin rates, it is not true that the 
``initial" kinetic energy of this shed core spin  
would necessarily find itself in the supernova.  Viscous 
dissipation conserves angular momentum and total energy,
but not kinetic energy.  A differentially rotating object
at a given angular momentum has free energy in its differential motion.
The rotational energy could end up in heat \citep{tqb}, which
in the protoneutron star and its inner envelope would be radiated in neutrinos.
Therefore, a large fraction of the 10$^{52}$ ergs of rotational energy
in a 5-millisecond protoneutron star could be converted into neutrinos
and would not have to power a super-energetic supernova.
This would be a small fraction of the total radiated neutrino energy
of a few $\times 10^{53}$ ergs.  Angular momentum is still conserved.

\section{Rotation-Induced Anisotropy of the Neutrino Field and Heating Profile}
\label{aniso}

For our core-collapse studies, we have constructed the first 2D 
multi-group, multi-angle, time-dependent radiation/hydrodynamics code 
in astrophysics.  In addition to being 6-dimensional (1(time) + 2(space) + 2(angles) + 1(energy groups)),
VULCAN/2D (Livne et al. 2004) has an ALE (Arbitrary Lagrangian-Eulerian) structure with remap, is axially-symmetric, 
can handle rotation, is flux-conservative, smoothly matches to the diffusion limit,
and is implicit in its Boltzmann solver, The implicit hydro 
variant has yet to be tested.  It also has a 2D MGFLD (multi-group, 
flux-limited diffusion) version that is computationally much
faster and allows us to more quickly explore parameter 
space.  What the code does not yet have are the
velocity-dependent terms in the transport equation, such as the Doppler shift
and aberration, though it does have the advection term.  It also does not have energy
redistribution due to inelastic scattering, though this will be incorporated in explicit fashion
(Thompson, Burrows, and Pinto 2003) in a subsequent version. 

Here, we present some stills from a preliminary 2D MGFLD calculation of the effect of rotation on the neutrino
radiation field.  Sixteen energy groups were used in these early test calculations.
In 2D, we simulated collapse, bounce, neutrino shock breakout, 
and the neutrino-driven convection stages for a 15 M$_{\odot}$ progenitor \citep{woosley}.  
Movies of these simulations, in particular detailing the evolution of the neutrino
flux vectors for various energy groups, are available from the first author upon request.

The results are summarized in Figures \ref{fig:3}, \ref{fig:4}, and \ref{fig:5}.  Figure \ref{fig:3} is a color map 
at 9.6 milliseconds after bounce of the entropy distribution, with the flux vectors for
electron neutrinos at 2.5 MeV superposed.  Note that the vectors are longer along the poles
than at the equator, reflecting the anisotropy of the neutrino emission induced by rotation.
Figure \ref{fig:4} shows the same quantities, but with
the flux vectors for the 13.9 MeV $\nu_e$ energy group superposed.  As is clear from
a comparison of Figures \ref{fig:3} and \ref{fig:4}, the fluxes and flux anisotropy are functions
of neutrino group.  Furthermore, and importantly, both figures demonstrate that
the entropy due to neutrino heating is larger along the poles than the equator,
qualitatively verifying one of the effects of rotation discussed in \S\ref{rotation}, 
and in Shimizu et al. (2001), Madokoro,~Shimizu,~\&~Motizuki (2004), and 
Kotake, Yamada, \& Sato (2003), that could aid explosion. Not shown is the anisotropy of the neutrino energy density 
contours that are oblate in the interior, but prolate outside near the shock wave.  
The rotationally-induced oblateness of the matter in the inner 
core can easily be seen in Figures \ref{fig:3} and \ref{fig:4}. 
This oblateness of the neutrino energy density in the interior is a 
consequence of von Zeipel's theorem concerning the coincidence of iso-effective-temperature
surfaces and isopotential surfaces.  Rotation makes the ``photosphere" oblate.  The  
prolateness of the neutrino energy density further out is related 
to the larger angle subtended by the core at the pole than the equator.
These are the first consistent time-dependent calculations of this effect, though
its basics have been investigated previously \citep{1monch89,monch89,shimizu,kotake,madokoro}.
Moreover, rotation has modified the convective plumes and created near bounce a barrel-shaped
structure rotating on cylinders. Figure \ref{fig:5} shows a map of Y$_e$ at 22.6 milliseconds
after bounce, with the 7.8 MeV $\nu_{e}$ flux vectors superposed.  At 
this early stage, for this calculation, the anisotropies of the 
Y$_e$ and electron-capture-rate distributions are dramatic.

Soon, we should know the true effect of a given degree of rotation on the supernova itself.
Whether the requisite rotation is provided by progenitor evolution has yet to be determined.

\section{Closing Remarks}
\label{conclusions}

The theoretical study of supernova explosions
is starting to couple multi-dimensional effects, neutrino radiation,
rotational effects, B-fields, magnetars, pulsar spins and kicks,
GRBs, hypernovae, MHD, and observed supernova blast morphologies
to obtain a consistent synthesis.  That synthesis has
not yet been accomplished, but its outlines are sharpening.
In the next year, we should be able to determine the role of rotation
and multi-dimensional transport effects on the supernova mechanism 
itself and answer many of the questions posed in this summary review.

\acknowledgments

We acknowledge discussions with
Todd Thompson, Jeremiah Murphy, Casey Meakin, 
Itamar Lichtenstadt, and Moath Jarrah.
Support for this work is provided in part by
the Scientific Discovery through Advanced Computing (SciDAC) program
of the DOE, grant number DE-FC02-01ER41184.  In addition, we thank
Jeff Fookson and Neal Lauver of the Steward Computer Support Group
for their invaluable help with the local Beowulf cluster and acknowledge
the use of the NERSC/LBNL/seaborg and ORNL/CCS/cheetah machines.

\section{Figure Captions:}
\label{captions}
\begin{figure}[!ht]
\caption{A schematic of the stalled shock wave after bounce, indicating the various important
regions.  They include the gain region, the cooling region, and the shocked region.  
Shown are the neutrino luminosity emerging from the core, the shock wave, and mass accretion. 
Note that the gain region is also the unstable region.  See text for a discussion.}
\label{fig:1}
\end{figure}

\begin{figure}
\caption{A snapshot of a rapidly rotating collapse and bounce simulation in 2D,
rendered in 3D with nested layers of isodensity contours from 10$^8$ gm cm$^{-3}$
to 10$^{13}$ gm cm$^{-3}$.  The funnel along the poles due to the
centrifugal barrier created after bounce by the rotation of the collapsing
Chandrasekhar core is clearly seen. Taken from Burrows, Ott, and Meakin (2003).}
\label{fig:2}
\end{figure}

\begin{figure}
\caption{A snapshot at 9.6 milliseconds after bounce  
of a simulation of the core of a rotating 11 M$_{\odot}$ progenitor.  The color map 
is of entropy and the vectors are flux vectors for the 2.5-MeV $\nu_e${s}.
See text for a discussion.}
\label{fig:3}
\end{figure}

\begin{figure}
\caption{The same as Figure \ref{fig:3}, but for the 13.9 MeV $\nu_e$ energy group.}
\label{fig:4}
\end{figure}

\begin{figure}
\caption{A color map of the Y$_e$ distribution 22.6 milliseconds after bounce.
Superposed are flux vectors for the 7.8 MeV $\nu_{e}$ energy group.  See 
text for a short discussion.} 
\label{fig:5}
\end{figure}


\begin{thebibliography}{}


\bibitem[Akiyama et al. 2003]{akiyama}
Akiyama, S., Wheeler, J.C., Meier, D., \& Lichtenstadt, I.,
2003, \apj, 584, 954

\bibitem[Bethe \& Wilson 1985]{bethe}
Bethe, H. \& Wilson, J.~R.~1985, \apj, 295, 14

\bibitem[Buras et al. 2003]{buras2003} Buras, R., Rampp, M., Janka, H.-Th., \& Kifonidis, K. 2003,
\prl, 90, 241101

\bibitem[Burrows,~Hayes,~\&~Fryxell 1995]{bhf}
Burrows, A., Hayes, J., \& Fryxell, B.A.~1995, \apj, 450, 830

\bibitem[Burrows \& Hayes 1996]{bh} Burrows, a. \& Hayes, J. 1996, \prl, 76, 352

\bibitem[Burrows \etal 2000]{B2} Burrows, A., Young, T., Pinto, P., Eastman,
        R. \& Thompson, T. 2000, \apj, 539, 865

\bibitem[Burrows \& Thompson 2002]{twilight}
Burrows, A. \& Thompson, T.A., 2002.
``The Mechanism of Core-Collapse Supernova Explosions: A Status Report,"
in the proceedings of the ESO/MPA/MPE Workshop
{\it From Twilight to Highlight: The Physics of
Supernovae}, p. 53, eds. Bruno Leibundgut and Wolfgang Hillebrandt (Springer-Verlag).

\bibitem[Burrows,~Ott,~\&~Meakin 2003]{bom} Burrows, A., Ott, C.D., \& Meakin, C. 2003,
to be published in the proceedings of ``3-D Signatures in Stellar Explosions:
A Workshop honoring J. Craig Wheeler's 60th birthday," held June 10-13, 2003, Austin, Texas, USA

\bibitem[Fryer \& Heger 2000]{fryer2000} Fryer, C.L. \& Heger, A. 2000, \apj, 541, 1033

\bibitem[Fryer \& Warren 2002]{fryer2002} Fryer, C.L. \& Warren, M. 2002, \apj, 574, L65

\bibitem[Fryer \& Warren 2004]{fryer2004} Fryer, C.L. \& Warren, M. 2004, \apj, 601, 391

\bibitem[Heger,~Langer,~\&~Woosley 2000]{heger00}
{Heger}, A., {Langer}, N., and {Woosley}, S.E. 2000, \apj, 528, 368

\bibitem[Heger,~Woosley,~\&~Langer 2003]{langer}
Heger, A., Woosley, S.E., \& Langer, N. 2003, in ``A Massive Star Odyssey: 
From Main Sequence to Supernova," Proceedings of IAU Symposium \#212, 
held 24-28 June 2001 in Lanzarote, Canary Islands, Spain. 
Edited by Karel van der Hucht, Artemio Herrero, and C\'{e}sar Esteban (San Francisco: 
Astronomical Society of the Pacific), p.357

\bibitem[Herant et al. 1994]{herant}
Herant, M., Benz, W., Hix, W.R., Fryer, C.L., \& Colgate, S.A. 1994, \apj, 435, 339

\bibitem[Hirsch,~Meynet,~\&~Maeder 2004]{hirschi} Hirschi, R, Meynet, G., \& Maeder, A. 2004,
submitted to \aa\ (astro-ph/0406552)

\bibitem[H\"oflich,~Wheeler,~\&~Wang 1999]{hoflich} H\"oflich, P., Wheeler, J.C., \& Wang, L. 1999, \apj, 521, 
179 

\bibitem[Hwang et al. 2001]{hwang01} Hwang, U., Szymkowiak, A.E., Petre, R., \& Holt, S. 2001,
\apj, 560, L175

\bibitem[Hwang et al. 2004]{hwang04} Hwang, U. et al. 2004, \apj, in press.

\bibitem[Janka \& M\"onchmeyer 1989a]{1monch89} Janka, H.-T. \& M\"onchmeyer 1989a, Astron. \& Astrophys., 209, L5

\bibitem[Janka \& M\"onchmeyer 1989b]{monch89} Janka, H.-T. \& M\"onchmeyer 1989b, Astron. \& Astrophys., 226, 69

\bibitem[Kotake et al. 2003]{kotake} Kotake, K., Yamada, S., \& Sato, K. 2003, \apj, 595, 304

\bibitem[Liebend\"{o}rfer et al. 2001]{lieben2001}
Liebend\"{o}rfer, M., Mezzacappa, A., Thielemann, F.-K., Messer,
O. E. B., Hix, W.~R., \& Bruenn, S.W.~2001, \prd, 63, 103004

\bibitem[Livne et al. 2004]{livne04}Livne, E., Burrows, A., Walder, R.,
Thompson, T.A., and Lichtenstadt, I. 2004, \apj, 609, 277

\bibitem[MacFadyen \& Woosley 1999]{macfadyen} MacFadyen, A.I. \& Woosley, S.E. 1999, \apj, 524, 262

\bibitem[Madokoro,~Shimizu,~\&~Motizuki 2004]{madokoro} Madokoro, H., 
Shimizu, T., \& Motizuki, Y. 2004, astro-ph/0312624

\bibitem[Maeda et al. 2002]{maeda} Maeda, K. et al. 2002, \apj, 565, 405 

\bibitem[Mazzali et al. 2003]{mazzali} Mazzali, P.A. 2003, \apj, 599, L95 

\bibitem[Mezzacappa et al. 2001]{mezz2001}
Mezzacappa, A., Liebend\"{o}rfer, M., Messer, O.E.B.,
Hix, W.R., Thielemann, F.-K., \& Bruenn, S.W.~2001, \prl, 86, 1935

\bibitem[Ott et al. 2004]{ott} Ott, C.D., Burrows, A., Livne, E., \& Walder, R. 2004,
\apj, 600, 834

\bibitem[Podsiadlowski et al. 2004]{pod} Podsiadlowski, P., Mazzali, 
P.A., Nomoto, K., Lazzati, D., \& Cappellaro, E. 2004, \apj, 607, L17 

\bibitem[Rampp \& Janka 2002]{rampp20022}
Rampp, M. \& Janka, H.-Th. 2002, \aa, 396, 331 

\bibitem[Shimizu et al. 2001]{shimizu} Shimizu, T., Ebisuzaki, T., Sato, K., \& Yamada, S. 2001, \apj, 552, 756

\bibitem[Scheck et al. 2004]{scheck}
Scheck, L., Plewa, T., Janka, H.-Th., Kifonidis, K., \& M\"uller, E. 2004, \prl, 92, 011103 

\bibitem[Thompson \& Duncan 1993]{duncan} Thompson, C. \& Duncan, R.C. 1993, \apj, 408, 194

\bibitem[Thompson,~Burrows,~\&~Pinto 2003]{Tod1} Thompson, T.A., Burrows, A., \& Pinto, P.A., 2003, \apj, 592, 434

\bibitem[Thompson,~Chang,~\&~Quataert 2004]{Tod2} Thompson, T.A., Chang, P., \& Quataert, E. 2004, \apj, 611, 380

\bibitem[TQB]{tqb} Thompson, T.A., Quataert, E., \& Burrows, A. 2004,
submitted to \apj, astro-ph/0403224 (TQB)

\bibitem[Thorstensen,~Fesen,~\&~van~den~Bergh 2001]{thor} Thorstensen, J.R., Fesen, R.A., \& van den Bergh, S.
2001, \aj, 122, 297

\bibitem[Wang et al. 2002]{wang1} 
Wang, L., et al. 2002, \apj, 579, 671 

\bibitem[Wang et al. 2003]{wang2} 
Wang, L., Baade, D., H\"oflich, P., \& Wheeler, J.C. 2003, \apj, 592, 457 

\bibitem[Willingale et al. 2002]{will02} Willingale, R., Bleeker, J.A.M., van der Heyden, K.J.,
Kaastra, J.S., \& Vink, J. 2002, \aa, 381, 1039

\bibitem[Willingale et al. 2003]{will03} Willingale, R., Bleeker, J.A.M., van der Heyden, K.J.,
\& Kaastra, J.S. 2003, \aa, 398, 1021

\bibitem[Woosley \& Weaver 1995]{woosley}
Woosley, S.E. \& Weaver, T.A. 1995, \apjs, 101, 181

\end{thebibliography}
\end{document}